\documentclass[acmsmall,screen,nonacm]{acmart}
\renewcommand\footnotetextcopyrightpermission[1]{}
\AtBeginDocument{%
  }

\usepackage{listings}
\usepackage{src/misc/llvm}
\usepackage{src/misc/nasm}
\usepackage{src/misc/c}
\usepackage{multirow}
\usepackage{tabularx}
\usepackage{verbdef}
\usepackage{booktabs}
\usepackage{subcaption}   % ← provides the 'subfigure' environment
\usepackage{diagbox} 
\usepackage{cleveref}
\usepackage{stfloats}
\usepackage[htt]{hyphenat}

\usepackage{tikz}
\usetikzlibrary{matrix,positioning}

\definecolor{greenY}{RGB}{166,247,166}
\definecolor{redN}{RGB}{255,38,0}
\definecolor{grayHeader}{RGB}{220,220,220}

\tikzset{ 
table/.style={
  matrix of nodes,
  row sep=-\pgflinewidth,
  column sep=-\pgflinewidth,
  nodes={rectangle,draw=black,text width=1.5ex,align=left},
  text depth=0.25ex,
  text height=1ex,
  nodes in empty cells
  },
texto/.style={font=\footnotesize\sffamily},
title/.style={font=\small\sffamily}
}

\begin{document}

%\title[No Bit Left Behind]{No Bit Left Behind:\\Fully Static Recompilation and Cross-Compilation of Complete Executables via Brute-Force Binary Lifting}
% \title[No Bit Left Behind]{No Bit Left Behind: Using Brute-Force Lifting to Achieve Scalable Fully Static Security Retrofitting, Recompilation, and Cross-Compilation of Arbitrary Binary Executables}
\title[No Bit Left Behind]{No Bit Left Behind: Using Brute-Force Lifting to Achieve Fully Static Binary Recompilation and Cross-Compilation of Arbitrary Binary Executables}

\author{Tianjiao Huang}
\email{tianjih1@uci.edu}
\orcid{0009-0008-5017-534X}
\affiliation{%
  \institution{University of California, Irvine}
  \city{Irvine}
  \state{California}
  \country{USA}
}

\author{Po-An Chen}
\email{poanc@uci.edu}
\orcid{0009-0006-9859-7096}
\affiliation{%
  \institution{University of California, Irvine}
  \city{Irvine}
  \state{California}
  \country{USA}
}

\author{Nick Baron}
\email{njbaron@uci.edu}
\orcid{0009-0005-1629-0385}
\affiliation{%
  \institution{University of California, Irvine}
  \city{Irvine}
  \state{California}
  \country{USA}
}

\author{Michael Franz}
\email{franz@uci.edu}
% \orcid{0009-0008-5017-534X}
\affiliation{%
  \institution{University of California, Irvine}
  \city{Irvine}
  \state{California}
  \country{USA}
}

% Find all of the underfull boxes
\overfullrule=5pt

\renewcommand{\shortauthors}{Huang et al.}

\newcommand{\mirrorball}{\textsc{MirrorBall}}
\newcommand{\intel}{x86-64}
\newcommand{\arm}{AArch64}

\newcommand{\cpp}{C\nolinebreak\hspace{-.05em}\raisebox{.3ex}{\small\bf +}\nolinebreak\hspace{-.10em}\raisebox{.3ex}{\small\bf +}}

\begin{abstract}

Binary recompilation is a technique for operating directly on executable code. It promises to automate two important tasks: retrofitting security mitigations onto legacy binaries, and migrating binaries across instruction set architectures (ISAs). Yet today, there is no fully automated system that can reliably lift arbitrary binary executables to a compiler intermediate representation (IR) such as LLVM IR, or that can fully statically and reliably translate non-trivial binary executables from one ISA to another. The main underlying problem is that  recovering a program's control flow graph (CFG) statically is impossible in general: computed branches can jump to targets that cannot be determined without actually running the program. Existing systems resort to runtime fallback mechanisms, requiring a significant portion of the binary translation machinery to accompany the translated program on the target machine.

This article presents a fully static, whole-program binary lifting system requiring no runtime translation support on the target. Rather than attempting to distinguish code from data, we treat every byte offset as a potential branch target and lift the entire binary in a \emph{brute-force} manner, constructing a \emph{superset CFG} that conservatively contains all feasible control flows. Statically unresolvable computed branches are thereby reduced to lookups in a dispatch table that points to the corresponding translated control flow path. We have implemented this approach as a prototype binary recompiler from x86-64 binaries to LLVM IR, requiring no code/data heuristics. We validate it with a fully static cross-compilation to AArch64, achieved by reusing existing LLVM backends with no modification.

\end{abstract}

\begin{CCSXML}
<ccs2012>
   <concept>
       <concept_id>10011007.10011006.10011073</concept_id>
       <concept_desc>Software and its engineering~Software maintenance tools</concept_desc>
       <concept_significance>500</concept_significance>
       </concept>
   <concept>
       <concept_id>10002978.10003022.10003465</concept_id>
       <concept_desc>Security and privacy~Software reverse engineering</concept_desc>
       <concept_significance>500</concept_significance>
       </concept>
 </ccs2012>
\end{CCSXML}

\ccsdesc[500]{Software and its engineering~Software maintenance tools}
\ccsdesc[500]{Security and privacy~Software reverse engineering}

\keywords{binary recompilation, binary lifting, cross-recompilation, static analysis, reverse engineering}

\iffalse
\received{20 February 2007}
\received[revised]{12 March 2009}
\received[accepted]{5 June 2009}
\fi

\maketitle{}

\section{Introduction}

Much of the world's software infrastructure runs on legacy binary executables: programs that were often created a long time ago and can no longer be properly maintained. There are many possible reasons why organizations might get stuck with such legacy binaries. A common cause is often a deprecated or broken toolchain, which makes it impossible to regenerate a legacy program even if its source code has survived. Another frequent problem is the absence of proper source code archiving or versioning discipline, leading to situations with multiple alternative surviving source files with unreliable file creation timestamps. It is often unclear which, if any, was used to create the legacy binary, and choosing the wrong one may resurrect subtle bugs that had already been corrected in the existing binary.

Being stuck with legacy binaries leads to follow-on problems. For example, a legacy binary may require the use of legacy hardware architectures, creating a secondary ``lock-in''. More seriously, legacy software may contain vulnerabilities that cannot easily be mitigated through conventional means, and its maintainers have no short-term remedy when one is discovered, even when these vulnerabilities become known to an attacker. 

Modernizing such legacy binaries is often both costly and slow. It may involve rewriting the original software, possibly using new libraries or even programming languages, adapting it to new compiler toolchains and/or operating systems, and then exhaustively testing that the end result has the same functionality as the superseded legacy binary. No ``quick fix'' is usually available: if an attacker finds an exploitable vulnerability in the legacy binary, or if the underlying hardware platform goes out of production and can no longer be reliably sourced, it will often take substantial time before any replacement executable is available.

For all of these and additional reasons, there has long been an interest in using existing binary code as the input to code modernization, removing the dependency on source code and source-code toolchains. For example, \emph{binary rewriting}~\cite{luk2005pin,bruening2003infrastructure,egalito,dinesh2020retrowrite,2019Panchenko_bolt} has been used to retrofit mitigations against security vulnerabilities into existing binary executables, \emph{binary translation}~\cite{rosetta_sunsetted,prj_champ_rosetta2,prism, chen2026elevator} has been used to migrate code from one instruction set architecture (ISA) to another, and \emph{binary lifting}~\cite{anand2013secondwrite, mcsema, Yadavalli2019, revng, altinay2020binrec, wytiwyg, deshpande2024polynima} translates a binary executable directly into a compiler intermediate representation (IR), exposing it to the downstream compiler ecosystem. While previous work has yielded many original and valuable insights and techniques, there has so far been no fully automated system that can reliably convert arbitrary (stripped) binary executables to a modern compiler IR, which would then further enable automatic retrofitting of compiler-inserted vulnerability mitigations or fully automatic cross-compilation to a different processor ISA than the code originated in.

This paper presents \mirrorball{}, a static binary recompiler whose central technique is \emph{superset disassembly}: rather than classifying each byte as code or data, it treats every byte offset in an executable segment as a potentially valid instruction start and constructs a superset control-flow graph (CFG) that encodes every such interpretation simultaneously. The resulting CFG is an over-approximation; it contains every path a dynamic execution could take, along with many that no execution ever will. Because every potentially valid interpretation is already represented, the recompiler does not need heuristics for the two undecidable decisions at the heart of static disassembly: code-vs-data classification and indirect target identification. The absence of heuristics in the superset CFG construction allows the use of heuristics in the optimization stages of the pipeline while still guaranteeing the correctness of the final recompiled binary.

Brute-force disassembly alone is not new. Prior work has used superset-style techniques for binary rewriting~\cite{bauman2018superset}, where the goal is to produce an instrumented binary that still executes on the same ISA as the original. The contribution of this work is not the disassembly technique in isolation, but its use within a whole-program lifting pipeline that produces a fully static, standalone output artifact in LLVM IR that can be fed to unmodified LLVM backends and middle-end passes.

The costs of this approach are substantial, and this paper characterizes them rather than obscures them. The whole-program emulation and the superset construction produce large code-size expansion, and lifted binaries run several times slower than their native counterparts for same-ISA recompilation, with further slowdown for cross-ISA.

Among our key contributions are the following:
\begin{itemize}
    \item We present \mirrorball{}\footnote{A mirrorball is a sphere fragmented into thousands of small mirrors, each reflecting light from a different angle. Similarly, \mirrorball{} disassembles a binary at every byte offset, producing a superset CFG that captures every possible control flow path.},
    the first static binary recompiler that doesn't rely on heuristics to make determinations about control flow and code-vs-data. It allows the recompiled binary to execute correctly without runtime translation support.

    \item We demonstrate the practicality of our approach with a fully static \intel{}-to-\arm{} cross-compilation system: lifted IR is handed directly to the unmodified LLVM \arm{} backend to produce a self-contained \arm{} executable, with no runtime translation support required on the target.

    \item We provide a thorough characterization of the code-expansion and performance slowdown that our brute-force method costs, and argue from this data that these costs are large enough to make brute-force lifting, in its current form, an unlikely candidate for practical deployment. We report this as a deliberate negative result: it quantifies what a maximally conservative, heuristics-free static lifter costs in practice, giving the field a concrete reference point against which future, less conservative designs can be measured.

\end{itemize}

\section{Background}
\label{sec:background}

\subsection{Binary Recompilation}
\label{sec:bg:recompilation}
A \emph{binary recompiler} translates (lifts) an executable into a compiler intermediate representation (IR), optionally applies modifications to the IR, and lowers it back to a new executable that is \emph{functionally equivalent} to the original~\cite{liu2022lifter_sok, mcsema, dbill, altinay2020binrec, engelke2020instrew, remill, Yadavalli2019, Hasabnis2016LiftingML}. That is, on every input that the original executable accepts, the two must produce the same output. Binary recompilation is distinct from \emph{binary rewriting}, which applies the modifications directly to the binary~\cite{e9patch, chamith2017InstructionPruning, he2026chimera, armore, bauman2018superset, luk2005pin, dynamorio, scott2001strata}, and from \emph{dynamic binary translation}, which translates the machine instructions to a different architecture at runtime~\cite{bellard2005qemu, engelke2020instrew, mettig2023rosetta2_threat}. %\Cref{sec:related_work} surveys representative systems from each category and situates \mirrorball{} among them; we give only the categories here.

\subsection{Binary Disassembly and Control-Flow Graph Reconstruction}
\label{sec:bg:disassembly}

Modifying a binary executable requires relocating symbols to maintain the correct references throughout the modified binary. However, much of the relocation information is lost in the compilation process.
A recompiler must find all the basic blocks to lift all the necessary binary code to IR. In order to do this, a recompiler reconstructs the control-flow graph (CFG) from the disassembled code.

For every byte in the binary, a recompiler must decide whether it is part of an instruction and, if so, what the basic block's control-flow successors are. The aggregate of these decisions over all bytes in the binary forms the CFG. An omitted edge causes the modified binary to run correctly until control reaches that edge, at which point it fails.

On variable-length ISAs like x86-64, instruction boundaries are not self-evident: the same bytes decode to different instruction sequences depending on the starting offset. Indirect control-flow transfers make this more difficult because their targets might not be able to be determined statically. Compilers typically generate indirect control-flow transfers for \cpp{} virtual dispatch, callbacks, jump tables for switch statements, and pointer mangling in \texttt{glibc} to protect long-lived code pointers (e.g., \verb|longjmp|).

Complete binary disassembly and full control-flow graph reconstruction are undecidable in general~\cite{kim2022pointer_analysis, engel2024Decidability, verbeek2024VerifiablyLifting, priyadarshan2023safer_binary_instrumentation}. Current practical recompilers address this in a few ways. \emph{Static disassembly with heuristics} infers indirect targets from compiler-specific patterns that are toolchain- and version-dependent~\cite{ddisasm, shen2013llbt, revng, mcsema, wang2015uroboros}. To fill in the remaining gaps, \emph{dynamic disassembly}~\cite{altinay2020binrec, deshpande2024polynima} runs the program on given inputs to observe its actual control flow. The coverage of dynamic disassembly is bounded by the inputs used. When all of the above approaches are insufficient, to ensure correctness of the modified binary when an omitted edge is encountered, systems will employ a \emph{runtime fallback} to handle missing control-flow edges, requiring that the binary modification system be shipped along with the modified binary.

\subsection{Superset Disassembly}
\label{sec:bg:superset_disassembly}

To solve the problem of undecidability in binary disassembly and control-flow graph reconstruction, the text section is disassembled at every byte offset, producing a \emph{superset disassembly}~\cite{bauman2018superset} of the real instruction stream. From this, we construct a \emph{superset CFG} whose nodes are every basic block in the superset disassembly, with over-approximations for indirect control-flow edges that treats every possible control-flow destination as a valid successor. The true CFG is a subset of the superset CFG, and the rest is translated dead code that the program never reaches.

The cost is size: a superset CFG contains many basic blocks and edges that the program never visits. The benefit is that the recompiler no longer needs heuristics for the two decisions that are undecidable in general: classifying a byte as code or data, and resolving the target of an indirect control-flow transfer. In our system, every offset is decoded, and every indirect branch becomes a table lookup over the full superset CFG, so no branch target is ever guessed and no candidate block is ever dropped on suspicion of being data.

Later stages of the pipeline do consult imprecise or incomplete information, for example, ELF symbols to recover multi-block functions (\Cref{sec:augmentation:function_recovery}), but this information never decides whether a control-flow edge exists. It only selects which of several already-correct implementations of that edge is used. A wrong guess can make the recovered IR larger or a call site slower; it cannot remove a feasible edge from the CFG or admit a path the original program could not take. \Cref{sec:design:static_analysis,sec:design:augmentation} identify each place this information is used and state what happens when it is wrong or absent.

The engineering challenge of managing the size of the superset CFG and keeping the translated program's performance within reason occupies much of the later sections. \mirrorball{} is the first system to extend superset disassembly to a complete, fully static recompilation pipeline that produces standard LLVM IR usable by unmodified backends.

\subsection{LLVM IR as a Lifting Target}

LLVM~\cite{lattner2004llvm} is an open-source, modular, and reusable compiler infrastructure, built around LLVM IR, a Single Static Assignment (SSA) IR, to allow different frontends and backends to work with the same intermediate representation. Lifting to the LLVM IR instead of to a custom IR gives the recompiled program access to dozens of unmodified backends and the mid-level analysis and optimization passes, and is not uncommon to use as an IR for binary lifting~\cite{engelke2020instrew, remill, Yadavalli2019, shen2013llbt}.

This choice has consequences for the rest of the system, LLVM IR was designed for frontends with access to source code; producing it from a binary requires encoding low-level machine state in a form the IR can represent. One machine instruction may have multiple effects on the program's state, and these effects must be captured in the IR. The registers are written to and read from multiple times in each instruction, and modeling them inefficiently can lead to high overhead in the generated code. The following sections discuss our solution to these challenges.

\section{Design Overview}
\label{sec:design}
\mirrorball{} is a fully static recompilation pipeline that lifts an executable binary into LLVM IR using brute-force lifting and then reconstructs a recompilable program from that IR. Since static analysis cannot determine which bytes of a binary are code, and since indirect control flow cannot be solved statically, we adopt a superset disassembly approach. Every byte offset in the executable sections is treated as a potential instruction boundary, and all control-flow transfers in these candidate instructions are incorporated into an over-approximated CFG. The recompilable IR is further optimized and instrumented before being compiled into a functionally equivalent binary.

\begin{figure*}
    \centering
    \includegraphics[width=\linewidth]{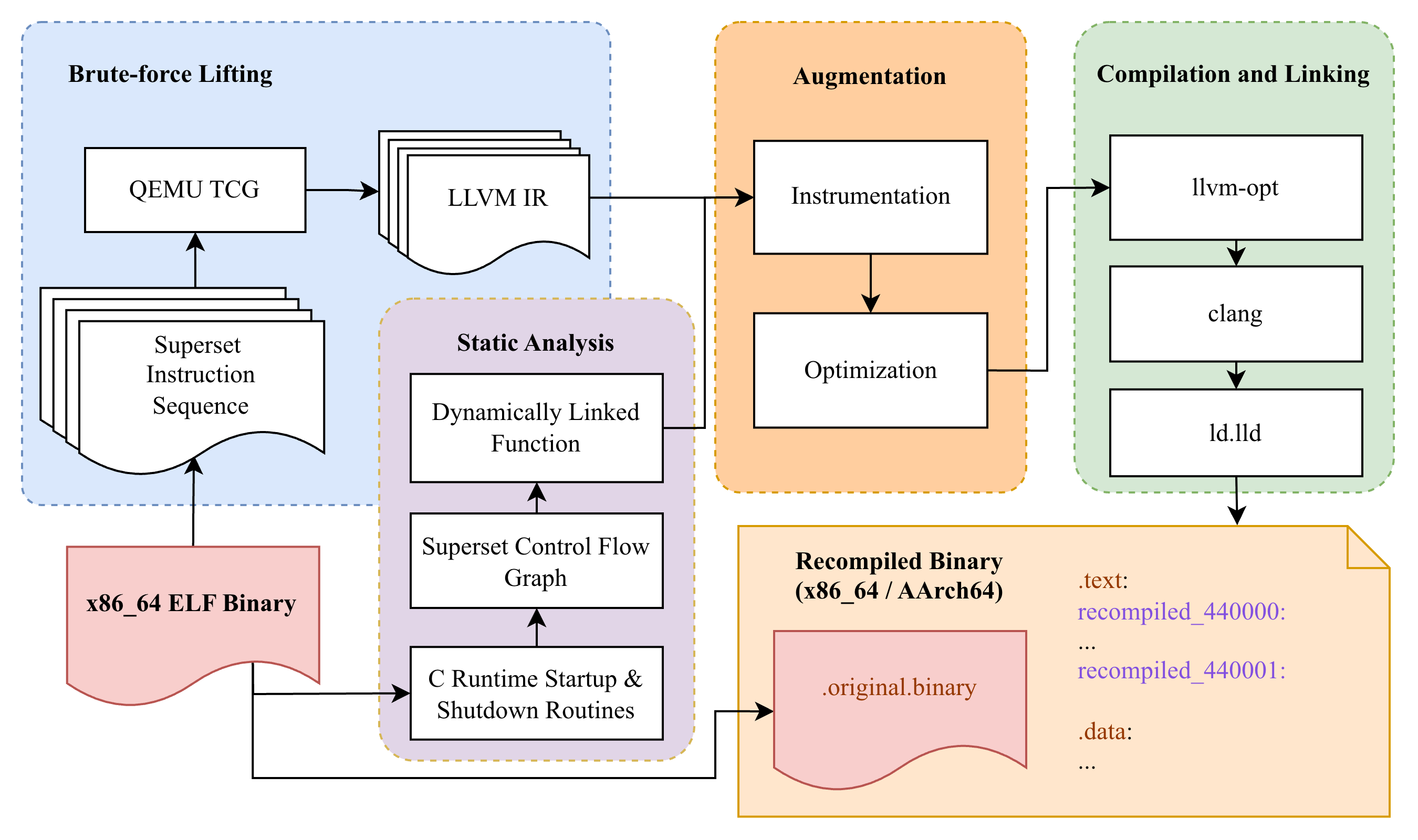}
    \Description{A visual summary of the architecture of \mirrorball{} }
    \caption{High-level architecture: After brute-force lifting to LLVM IR, we lower back either to an \intel{} binary or to a cross-compiled \arm{} binary. In the output binary, each lifted basic block becomes a function named after its offset in the input binary (e.g., \texttt{recompiled\_440000} for the block at offset \texttt{0x440000}); the input binary itself is embedded verbatim as a read-only data blob, labeled \texttt{.original.binary}, so that data references into it (\Cref{sec:design:compilation_linking}) resolve correctly.}
    \label{fig:architecture}
\end{figure*}

As presented in \Cref{fig:architecture}, our recompilation pipeline consists of the following stages: 
\begin{enumerate}
    \item \textbf{Brute-Force Lifting:} The executable sections of the input binary are disassembled at every byte offset. Each offset with a valid instruction is treated as the start of a basic block, which is lifted to LLVM IR. At this stage, each basic block is a function in its own right, with no direct control flow entering or exiting. (\Cref{sec:design:brute_force_lifting})
    \item \textbf{Static Analysis:} This stage parses the ELF file to recover the superset CFG, the dynamically linked functions, and other runtime bootstrapping information. (\Cref{sec:design:static_analysis})
    \item \textbf{Augmentation:} Taking the results from the previous stages, this stage augments the LLVM IR with additional information and optimizations to allow it to be consumed by unmodified backends. (\Cref{sec:design:augmentation})
    \item \textbf{Compilation and Linking:} The final IR is compiled and linked against the input binary to produce a functionally equivalent executable. (\Cref{sec:design:compilation_linking})
\end{enumerate}
%
% Figure~\ref{fig:in_n_out} presents a simplified human-readable sketch of the transformations that \mirrorball{} applies. 
% Block (a) shows an example input program that is fed into the system in binary form.
% Block (b) presents a recovered and recompilable LLVM IR snippet of \verb|push rbp| highlighted in block (a), where typed pointers and direct manipulation of architectural state are visible, enabling code generation from the IR as written.
% Block (c) shows the output program corresponding to the input program, generated from the LLVM IR.

\verbdef{\astar}{astar}
\verbdef{\bzip2}{bzip2}
\verbdef{\gcc}{gcc}
\verbdef{\gobmk}{gobmk}
\verbdef{\h264ref}{h264ref}
\verbdef{\hmmer}{hmmer}      
\verbdef{\libquantum}{libquantum} 
\verbdef{\mcf}{mcf}
\verbdef{\omnetpp}{omnetpp}
\verbdef{\perlbench}{perlbench}  
\verbdef{\sjeng}{sjeng}      
\verbdef{\xalancbmk}{xalancbmk}  

\section{Brute-Force Lifting}
\label{sec:design:brute_force_lifting}

The first stage of \mirrorball{} produces LLVM IR for the input binary's executable code, before any control flow between blocks has been reconstructed. Because the precise basic blocks of a binary cannot be determined statically~\cite{bauman2018superset}, we do not commit to a single decoded view of the program. Instead, we start decoding at every byte offset of every executable section, and treat every offset that begins a valid instruction sequence as a potential basic block entry.

When decoding a given offset, we continue instruction by instruction until a basic block terminator is reached. A basic block is a sequence of instructions with a single entry point and a single exit point. We treat \verb|jmp|, \verb|call|, and \verb|ret| as terminators. \verb|call| is a terminator because the callee may not return to its caller. \verb|ret| is a terminator and an indirect transfer as the return address stored in the stack may have been modified. If the decoder encounters an invalid opcode or a privileged instruction, the offset is discarded.

We lift each surviving candidate block in isolation using a QEMU-based frontend, building on the lifting frontend of the publicly available S2E project~\cite{chipounov2011s2e}. It lifts each individual block into QEMU TCG IR, an IR intended to help the emulator to generate code for different architectures, before lifting it to LLVM IR. Lifting the complex instructions, such as the \texttt{rep} extension and vector instructions, are handled transparently by the QEMU frontend. The result is one LLVM IR function per candidate basic block, with no control flow crossing function boundaries. Architectural state that does not fit LLVM's SSA model (i.e., the general-purpose registers, the program counter, the \verb|EFLAGS| register, the SIMD registers) is modeled as global variables, read and written by each lifted function as needed. The lifted stack is held in a separate region of emulated memory, separated from the host's native stack to hold the register spills from the input program's execution.

We chose QEMU's instruction semantics over an LLVM-IR lifter such as McSema~\cite{mcsema} or Remill~\cite{remill} for two reasons. First, QEMU's x86-64 decoder and semantics are exercised continuously by its use as a general-purpose emulator, giving us production-tested coverage of instruction classes, such as the \texttt{rep} family and the vector extensions, that we did not have to validate ourselves. Second, and more directly relevant to this paper's central claim, McSema and similar LLVM-IR lifters take an externally supplied CFG as input, typically recovered by a disassembler such as IDA Pro; the correctness of the lifted program then inherits whatever code/data and indirect-target heuristics that external tool used to build the CFG. Since removing exactly that dependency is the contribution of this work, adopting such a frontend as-is would have reintroduced the heuristic we set out to eliminate, and repurposing one to consume a superset CFG instead of its own recovered CFG would have been a comparably large undertaking to building the brute-force pipeline described here. This choice is not without cost: the register-as-globals model that QEMU's frontend leaves us with is, by our own measurement (\Cref{sec:eval:lifter_overhead}), the single largest source of the runtime overhead we report, and we do not consider that cost fully settled by this design choice.

At the end of this stage the input program is represented as a collection of single-block LLVM IR functions. The collection is valid LLVM IR, but without the control flow information that connects the basic blocks.

% --------------------------------------------------------------------
\section{Static Analysis}
\label{sec:design:static_analysis}
The second stage parses the brute-force lifted sequence and the input binary's ELF metadata to recover the information needed to connect the lifted blocks and to handle the program's interactions with its runtime environment. This stage extracts the necessary information for the later stages to reconstruct a functionally equivalent binary.

\paragraph{Superset CFG edges.}
For each candidate block, we identify its successor from its terminator instruction.
Direct branches have one or two constant successors, while indirect ones, including every \verb|ret| instruction, may target any valid basic block.
The resulting graph is the \emph{superset CFG}: an over-approximation that contains every feasible control-flow edge, along with a much larger number of infeasible ones.
The augmentation stage uses this graph to install inter-block control-flow. 
The infeasible edges contribute compile-time and code size overhead but have no functional impact, because the recompiled binary can only follow edges that the program actually takes.

\paragraph{Symbol-based function hints.}
The ELF symbol table marks the entries of named functions in the input binary.
Starting from each symbol that points into the \verb|.text| section, we walk forward through statically resolvable direct edges by following direct jumps and stopping at any indirect transfer.
The blocks reached in this walk are flagged as the likely intra-procedural body of a named function.
Blocks that are not reached, which include most of the superset, remain unflagged.
The augmentation stage uses this information to decide which blocks to fold into recovered functions and which to leave as single-block functions (\Cref{sec:augmentation:function_recovery}).

\paragraph{Dynamically linked function signatures.}
An ELF binary might rely on external functions residing in libraries that are dynamically linked at runtime. In this paper, we refer to them as external functions or dynamically linked functions. The \verb|.dynsym| section lists the symbols that the dynamic linker will resolve at runtime. For \cpp{} symbols, the Itanium \cpp{} ABI's mangling scheme~\cite{itanium_abi} encodes the full function signature, including the parameter list. The mangled symbol \verb|_ZNSo5writeEPKcl|, for example, decodes to \verb|std::ostream::write(char const*, long)|; counting the implicit \verb|this| parameter, this is an effective signature of \texttt{(size\_t, size\_t, long)}.

C symbols only contain their symbol names and do not encode their parameter lists. For these, we maintain a per-symbol mapping populated from the documentation in their \textit{man} pages. This mapping is exact information, not a heuristic guess, but it is incomplete by construction: the current prototype covers only the C/\cpp{} library functions called by our evaluation suite. A call whose callee is not covered by this mapping and cannot be resolved through \verb|.dynsym| mangling is instead marshalled by the generic external-call trampoline described in \autoref{sec:augmentation:external_calls}; consulting a recovered signature first is purely a performance optimization that avoids that trampoline's more conservative, and slower, calling sequence.

\paragraph{Bootstrapping metadata.}
The analysis records the ELF entry point, the contents of \texttt{.init\_array} and \texttt{.fini\_array}, and the addresses that the input binary references through the global offset table.
These are needed during compilation and linking (\Cref{sec:design:compilation_linking}) to reproduce the C runtime library's startup and shutdown behavior.

\section{Augmentation}
\label{sec:design:augmentation}
After lifting and analysis, the IR is a set of disconnected single-block functions annotated with edge information, function hints, and external signatures. The augmentation stage installs the CFG, recovers functions that are more likely to be executed, and inserts trampolines required at the boundary with dynamically linked libraries. After this stage, the IR is accepted by the unmodified LLVM pipeline.

\subsection{Control-Flow Patching}
\label{sec:design:control_flow_patching}
\paragraph{Direct branches.}
For a direct branch, the target offset is a compile-time constant. Since each lifted block resides in its own LLVM function, we emit a call to the lifted function corresponding to that offset. If the offset is invalid, we emit a call to \verb|abort| instead, which is unreachable in any correct execution and is removed by later optimization passes.

\paragraph{Indirect branches and returns.}
For indirect branches, the target is computed at runtime. Every indirect transfer is routed through the \verb|dispatch_indirect| function, which takes the indirect target, specified by the value of the emulated program counter, and transfers control to the lifted function corresponding to that offset. Offsets with invalid basic blocks are handled by emitting a call to \verb|abort|. Offsets that fall in the Procedure Linkage Table are from indirect calls to dynamically linked functions and are routed onward to the external call trampoline (Subsection~\ref{sec:augmentation:external_calls}) instead.

\verb|ret| instructions are lowered the same way. Because the return address lives in emulated stack memory that the callee may have written to, we cannot lower \verb|ret| as a native LLVM return; the indirect dispatcher is the only correct lowering.

%\tian{Potential cut?} \nick{do it}
%A straightforward implementation of \verb|dispatch_indirect| is a switch statement whose cases enumerate the valid offsets and let the LLVM perform the appropriate optimizations. However, programs of any realistic size have at least tens of thousands of superset basic blocks, which increases the compilation time and hurts the runtime performance. We implemented \verb|dispatch_indirect| as a bound-checked array indexed by the emulated program counter. LLVM is then able to inline the dispatcher at every call site. On \mcf{}, the array-based dispatcher reduces indirect dispatch overhead by 20\% relative to the switch based version.

\paragraph{Tail-call optimization.}
Because returns are lifted as indirect calls rather than as native returns, a lifted function never truly returns to its caller.
It instead tail-calls into the block pointed to by the return address.
Without further intervention, this means the native stack grows for the entire lifetime of the lifted program.
We avoid this by giving every lifted function a uniform calling convention and marking every call site as \verb|musttail|.
LLVM then lowers these as jumps, and the native stack remains bounded by the deepest non-tail call sequence in the runtime support code.

\subsection{Argument for CFG Completeness}
\label{sec:augmentation:cfg_completeness}
The central correctness claim of this paper is that the superset CFG contains every control-flow edge a real execution of the input program can take, for every input, not just the ones exercised by our evaluation suite. This subsection states the argument for that claim explicitly, and its boundary: what it establishes, and what it deliberately leaves as a separate, narrower assumption.

\paragraph{Every real instruction-start offset survives lifting.}
A real execution can only ever fetch an instruction from a byte offset within a mapped executable section (excluding the runtime code generation we exclude by construction, \Cref{sec:future_work:self_modifying_code}). Brute-force lifting (\Cref{sec:design:brute_force_lifting}) decodes every byte offset in every executable section, with no code/data classification step that could skip one, so any offset a real execution could ever use as an instruction start is among the offsets we attempt to decode.

x86-64 decoding is a deterministic function of the byte sequence and the offset at which decoding begins: the same bytes starting at the same offset always decode to the same instruction, independent of what any other offset in the binary decodes to. If a real execution reaches offset $O$ and executes the instruction the hardware decodes there, then our decoder, decoding from that same offset $O$, decodes the identical instruction, because it is reading the identical bytes at the identical starting point. The one way a candidate block is discarded is an invalid opcode or a privileged instruction at its start (\Cref{sec:design:brute_force_lifting}); by the determinism argument, this can only happen at $O$ if the real hardware would also fault or trap there. Discarding such an offset therefore never discards a path a correct execution could take on any input; at worst, it changes how the failure is observed (an \verb|abort| in the lifted program rather than the original's own \verb|SIGILL|/\verb|SIGSEGV| delivery or system-call trap), which we note as a narrow deviation in observable behavior that is out of scope of this paper.

\paragraph{Every real edge is installed.}
Given that every real instruction-start offset survives as a lifted block, every control-flow edge a real execution takes must also be preserved. Direct edges are compile-time constants read from the instruction bytes and patched to a call to the corresponding lifted function (\Cref{sec:design:control_flow_patching}); by the same determinism argument, if a real execution takes that edge, the target offset is valid and was lifted, so the patched call is never the \verb|abort| fallback on that path. Indirect edges, including every \verb|ret|, are resolved at runtime through \verb|dispatch_indirect| over the full set of surviving offsets, with no further heuristic filtering of indirect targets (\Cref{sec:design:control_flow_patching}); whatever offset the real execution computes as an indirect target, if it is an offset a real execution could reach, it survived lifting by the preceding argument, and \verb|dispatch_indirect| finds it.

\paragraph{What this argument does not cover.}
This argument establishes that the superset CFG is complete with respect to control-flow edges internal to the lifted program. It does not, by itself, establish that every mechanism at the boundary of that CFG is correct. We rely on three further, separately stated assumptions: (i) that the QEMU-derived semantics for a decoded instruction faithfully reproduce its architectural effect, which we inherit from QEMU's maturity as a general-purpose emulator rather than re-derive; (ii) that the external-call marshalling described in \Cref{sec:augmentation:external_calls} correctly bridges the two ABIs for the call shapes it claims to support, with the composite-return limitation noted there as an explicit, scoped exception; and (iii) the exclusions already stated in \Cref{sec:design:brute_force_lifting} and the discussion of future work, namely no self-modifying code, no multithreading, and no \cpp{} exception unwinding in the current prototype. We consider (i) and the excluded input classes to be inherited or scoped limitations rather than gaps in the argument above, and (ii) to be the one place where an incorrect implementation, rather than an incomplete CFG, could still produce a wrong answer.

\subsection{Function Recovery}
\label{sec:augmentation:function_recovery}
If every basic block remains its own function, the lifted binary suffers three costs simultaneously. Every control flow edge lowers to a \verb|call|, which is expensive at runtime. Basic blocks that frequently jump between each other may get placed further apart, which adds pressure to the instruction cache and further hinders the performance. LLVM's function-level passes, such as \verb|mem2reg|, cannot optimize across basic block boundaries, suppressing most cross-block optimizations.

Naive merging is not viable, as a source basic block of $N$ bytes yields $N$ distinct superset basic blocks, each of which may be entered as function entry point independently. If $K$ of the other superset basic blocks are successors, merging each successor into each of its predecessor superset functions would produce on the order of $N \times K$ duplicated blocks per source block, and the cost compounds as recovery proceeds.

Without impacting the functional correctness of the lifted program, we opt to use heuristics to assist in recovering functions that start at the addresses that are likely to be function entry points in an actual execution. Although there are many other heuristics available for function recovery~\cite{shirani2017binshape, bao2014byteweight, calvet2012aligot, shin2015recognizing_functions_neural}, our prototype uses only the ELF symbol table information. For each ELF symbol pointing into the \verb|.text| section, we connect the blocks reachable from that symbol along statically resolvable edges into one LLVM IR function. The resulting IR contains a few large functions, embedded in a much larger collection of single-block functions that cover the rest of the superset. 

% \autoref{fig:function_recovery} contrasts the strategy with naive merging.

% \begin{figure*}
%     \centering
%     \includegraphics[width=1\linewidth]{src/fig/Function Recovery.png}
%     \caption{Function Recovery: Comparison of the function recovery strategies given the input binary. This figure assumes instructions are always one-byte to simplify the illustration.}
%     \label{fig:function_recovery}
% \end{figure*}

This process is approximate: a symbol-walk may include tail-calls, and indirect targets within a named function are not folded into it. But the approximation does not affect correctness, only performance, every block, recovered or not, remains reachable through \verb|dispatch_indirect|. %\Cref{sec:function_recovery_eval} reports the resulting reduction in compilation time and binary size.

\subsection{External Call Boundary}
\label{sec:augmentation:external_calls}
ELF binaries use Procedure Linkage Tables (PLT) to facilitate dynamic linking and resolved dynamically linked symbols. When a program tries to invoke a dynamically linked function, it goes through the PLT, which redirects the call to the actual function at runtime.

When control reaches a PLT entry, the lifted program crosses from emulated execution into native execution. On the emulated side architectural state lives in threadlocal variables and on the emulated stack; on the native side, the shared library expects its arguments in the host architecture's physical registers and on the host's stack. The augmentation stage inserts marshalling code at every such boundary.

The shape of that code depends on the calling conventions of both sides. Our prototype targets Linux on \intel{} and Linux on \arm{}, which follow the System V AMD64 ABI~\cite{system_v} and AAPCS64 respectively~\cite{aapcs64}. The two conventions agree in many places but differ in two ways that affect every external call: \intel{} dedicates six integer argument registers whereas \arm{} dedicates eight, and the two architectures handle composite return values differently. \autoref{tab:abi} summarizes these differences.

\begin{table}[tbp]
\centering
\begin{tabular}{l c c}
\toprule
\multicolumn{1}{c}{\textbf{ABI Features}} & \multicolumn{1}{c}{\textbf{\intel{}}}             & \multicolumn{1}{c}{\textbf{\arm{}}} \\
\midrule
  \multirow{2}{*}{Integer arguments 1-6}  & \verb|rdi|, \verb|rsi|, \verb|rdx|,   & \multirow{2}{*}{ \texttt{x0-x5} } \\
 & \verb|rcx|, \verb|r8|, \verb|r9| &  \\
  Integer arguments 7-8                  & Stack                         & \verb|x6-x7| \\
  Integer arguments 9+                   & Stack                         & Stack \\
  Floating-point args. 1-8           & \verb|xmm0-xmm7|    &  \verb|v0-v7|     \\
  Return value                          & \verb|rax|, \verb|rdx|                      & \verb|x0|, \verb|x1| \\
  Return address                        & Stack                         &  \verb|lr|, Stack     \\
  Stack alignment                       & 16 bytes                      & 16 bytes \\
\bottomrule
\end{tabular}
\caption{Overview of \intel{} and \arm{} calling conventions~\cite{system_v,aapcs64}. }
\label{tab:abi}
\end{table}

\paragraph{Calls with known signatures.}
When a callee's signature is recorded by the static analysis phase, the marshalling code is straightforward. We read the arguments from the emulated registers and stack according to the input binary's calling convention, place them in the parameter list, and emit a direct call. The LLVM backend will move the arguments to the native registers and stack according to the host's calling convention. This is the common case in our evaluation suite, where \verb|.dynsym| supplies the signatures of \cpp{} functions through name mangling and our manually transcribed mapping covers C functions invoked by the benchmarks.

\paragraph{Calls with unknown signatures.}
Some calls cannot be marshalled this way. Variadic functions are an example: the number and the type of arguments are determined by a format string or other convention, and might not be known at lifting time. For these, along with any functions with unknown signatures, we route through a generic trampoline:

\begin{enumerate}
    \item Copy the emulated argument registers to the corresponding native registers.
    \item Prepare a new stack containing the spilled arguments and point the native stack pointer to it.
    \item Point the emulated stack pointer to the native stack.
    \item Call the external function.
    \item Restore the stack pointers.
    \item Copy the return value from native registers into emulated state.
    \item Return to the caller.
\end{enumerate}

The trampoline does not know the precise argument count, so it conservatively copies the first six integer registers and the first eight floating registers. There is no consequence in populating argument registers that the callee does not actually read: a callee's compiled code only ever consumes the registers corresponding to its own parameter list, so values placed in registers beyond that list are simply never loaded. Nor does over-population corrupt the caller's state, since argument registers are caller-saved (call-clobbered) under both the System V AMD64 ABI and AAPCS64, meaning neither convention requires their contents to survive the call. On the return path, the trampoline copies the callee's scalar return registers back into emulated state; it does not currently marshal composite (larger-than-register) return values, whose caller-allocated return slot is addressed by a different register on each ABI (\verb|rdi| on \intel{}, \verb|x8| on \arm{}). This is a gap in the current prototype rather than a consequence of the approach: every C/\cpp{} library function invoked by our evaluation suite returns a scalar, and extending the trampoline to composite returns requires only bookkeeping, not a new mechanism.

\begin{figure}
    \centering
    \includegraphics[width=1\linewidth]{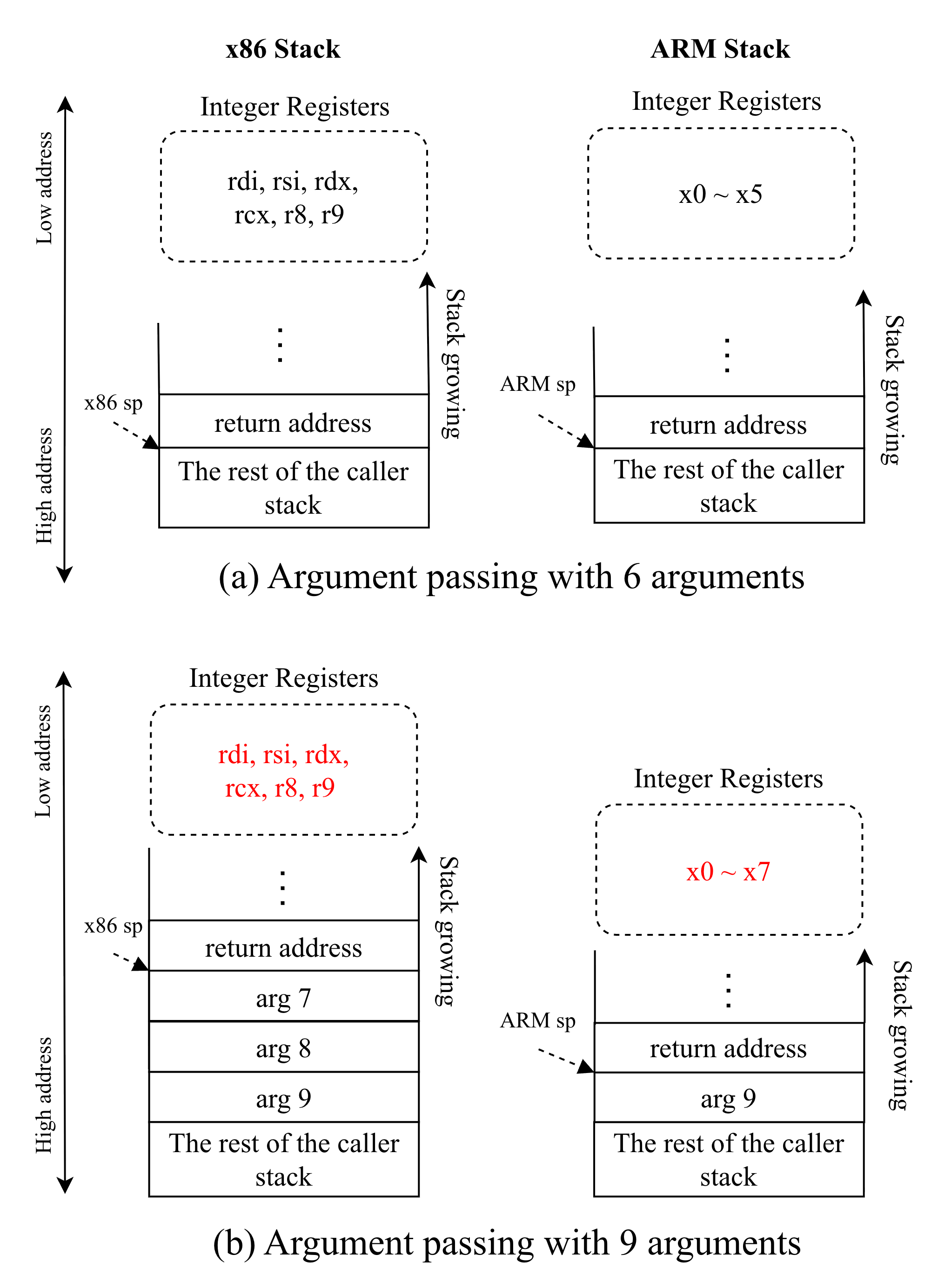}
    \Description{A diagram contrasting how the seventh and eighth integer arguments are placed on the stack under the x86-64 System V ABI versus kept in registers x6 and x7 under AAPCS64, with later arguments spilling to the stack on both architectures.}
    \caption{Argument spilling: On \intel{} and \arm{}, the first six integer arguments use registers. The seventh and eighth spill to stack on \intel{} but remain in a register on \arm{}. Beyond the eighth, all remaining integer arguments spill to the stack.}
    \label{fig:stack_spilling}
\end{figure}

\paragraph{Argument spilling.}
When the number of arguments exceeds the available argument registers, both ABIs spill arguments to the caller's stack in source order. Because \intel{} has two fewer integer argument registers than \arm{}, an integer argument in the seventh or eighth position resides on the emulated stack on \intel{} but in register \verb|x6| or register \verb|x7| on \arm{}. The trampoline reads these two slots from the emulated stack and writes them to \verb|x6| and \verb|x7|, keeping the remaining spills on a stack pointed to by the native stack pointer before calling the external function. \autoref{fig:stack_spilling} illustrates this.

% Floating pointer crisscrossing. \tian{too much details and cut this?}. \autoref{fig:interleave_stack_spilling}

% When floating-point arguments also spill, the layout of the spill
% region on \intel{} and on \arm{} can differ, because the two
% architectures spill different positions. Figure~\ref{fig:stack_spilling}
% shows the basic case; Figure~\ref{fig:interleave_stack_spilling} shows
% a harder one in which a floating-point spill appears before an
% integer spill in the argument list. In this case the spill region
% cannot be transformed by a single pointer adjustment, and the
% trampoline would need per-slot type information that an unidentified
% variadic call does not provide. We detect this condition at runtime
% using the System V convention that \verb|al| reports the number of
% floating-point arguments passed in vector registers; when this
% indicates that a floating-point spill is present, the trampoline
% halts and reports the call site so that a bespoke signature can be
% supplied. We have not encountered this case in our evaluation suite.

\paragraph{External stack management.}
\label{sec:augmentation:external_stack_management}
Since the emulated \intel{} environment has two fewer argument registers than the host \arm{} environment, a naive solution to allow the external functions to access the argument passed via the emulated stack is by pointing the native stack pointer to \verb|Emulated_RSP + 16|. The 16 bytes ``popped'' by this operation could be one of the following:

\begin{itemize}
    \item \textbf{Excess arguments spilled} onto the emulated stack if the argument registers are exhausted. 
    %And if the argument registers are not exhausted, these 16 bytes could contain one of the following:
    \item \textbf{Caller's immutable data} that the caller does not expect to be modified across function calls (e.g., callee-saved registers).
    \item \textbf{Caller's mutable data} that the caller expects to be modified across function calls. For example, the caller might pass a pointer to a buffer on the stack and expect the \verb|scanf| callee to write into it.
\end{itemize}

If the data is meant to be immutable, the trampoline could restore its value after the external call, or it could keep the modification if the data is meant to be mutable. Implementing this would require complex pointer analysis. Instead, we take a conservative approach by assuming these 16 bytes could be any one of the above. We construct a separate \emph{external stack} for the duration of each call. Pages from the current pointer up to the next page boundary are copied onto the external stack so that stack arguments remain addressable. Pages below this boundary are mapped as read-only and shared with the emulated stack, so that reads through caller pointers that fall there return the correct values. Pages above are allocated fresh as the callee needs them. After the call, the external stack is set aside and could be reused if similar mapping is needed for another call. %\autoref{fig:external_stack_management} illustrates this.

% \begin{figure*}[tbh]
%     \centering
%     \includegraphics[width=\linewidth]{src/fig/external_stack_management.png}
%     \caption{External stack handling: External calls may overwrite stack arguments or caller memory, making reuse of the native stack unsafe. To isolate these cases, a separate external stack is created by copying the active argument region, mapping the higher-address pages as shared read‑only with the emulated stack for efficient access to the stack arguments, and allocating new writable pages for lower addresses for normal stack growth.}
%     \label{fig:external_stack_management}
% \end{figure*}

% Tian: cut this since it's simple and not very interesting.
% \paragraph{Return values.}
% For scalar returns, the trampoline copies the native return register
% or registers back into emulated state. For composite returns that do
% not fit in registers, both ABIs require the caller to pass a pointer
% to a caller-allocated return slot; the register used to pass that
% pointer differs (\verb|rdi| on \intel{}, \verb|x8| on \arm{}). The
% current prototype omits support for composite returns from external
% functions, since the C library functions invoked by our benchmark
% suite return only scalar types.

\begin{figure*}[tbp]
    \centering
    \includegraphics[width=\linewidth]{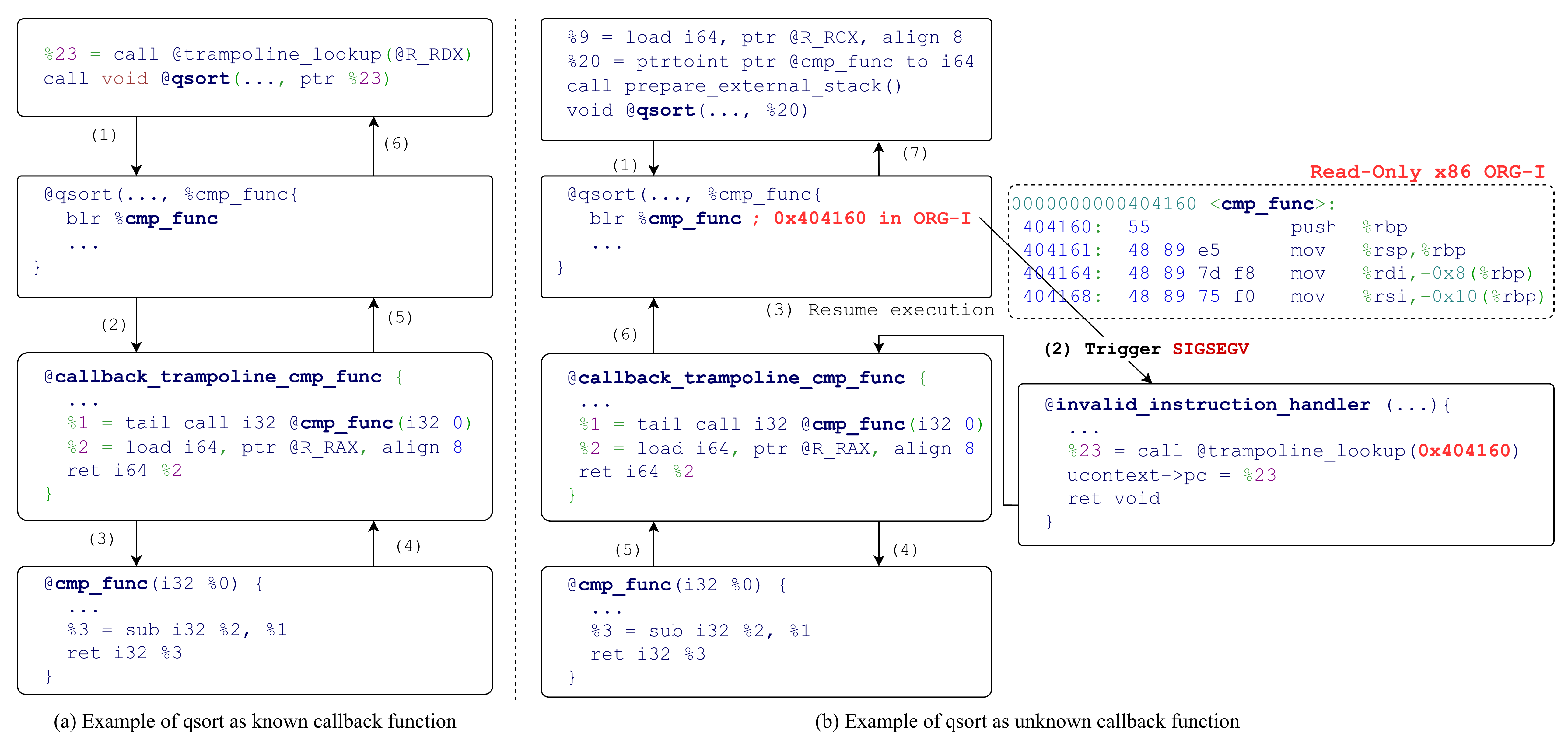}
    \Description{Two side-by-side call sequence diagrams for a qsort comparator callback. In the known-callback case (a), the call site is patched ahead of time to pass the address of a generated callback trampoline directly to qsort, which the trampoline uses to invoke the original lifted comparator and return its result. In the unknown-callback case (b), qsort is called with the original code-segment address unmodified; dereferencing that address raises a SIGSEGV because the segment is mapped non-executable, and a signal handler resolves the faulting address through trampoline\_lookup and resumes execution at the corresponding trampoline.}
    \caption{External Callback Handling: Illustrated steps and handling of replacing callback functions with callback trampolines. All IRs were extensively simplified for readability.}
    \label{fig:callback}
\end{figure*}

\subsubsection{Callbacks}
\label{sec:augmentation:callbacks}

Callbacks are function pointers passed via function parameters and may be invoked by the callee at a later time. Callbacks within the input program are handled by the \verb|dispatch_indirect| transparently since they are a form of indirect control flow, but callbacks that cross the boundary between the lifted program and the native libraries require special handling.

When a callback is passed from the lifted program to a native library, the pointer is an address in the input program's code segment, which on the cross-architecture side is a sequence of \intel{} instruction bytes; if the external library calls it directly, the host architecture will attempt to decode \intel{} bytes as \arm{} instructions and fault almost immediately.

For function-pointer arguments whose role is recorded in the static analysis phase (e.g., the comparator passed to \verb|qsort| or \verb|std::binary_search|), we replace the pointer at the call site with the address of a \emph{callback trampoline}. The trampoline is a small native function that demarshals arguments from native state into emulated state, calls the lifted function corresponding to the original address, and remarshals the return value back into native state. Since every byte offset in the input program could be a callback target, and we could not statically determine which ones are actually used as callbacks, we conservatively generate a trampoline for every superset basic block. Trampolines are looked up by a generated function \verb|callback_trampoline_lookup|, which maps the input binary addresses to trampoline addresses.

For function-pointer arguments whose role is not recorded, the call site cannot be patched, because we cannot identify which argument holds the function-pointer. Instead, we mark the input binary's code segment non-executable in the lifted binary. When the external library dereferences an unrecognized pointer into that segment, a \verb|SIGSEGV| signal is raised; our signal handler then invokes \verb|trampoline_lookup| on the faulting address and resumes execution at the corresponding trampoline. If the address falls outside the code segment, we can transfer control to the signal handler installed by the input program instead to handle its custom signal handling logic. \autoref{fig:callback} illustrates this.

% \begin{figure}
%     \centering
%     \includegraphics[width=0.5\linewidth]{src//fig/callback_trampoline_lookup.png}
%     \caption{Enter Caption}
%     \label{fig:placeholder}
% \end{figure}

\subsubsection{\texttt{setjmp} and \texttt{longjmp}}
\label{sec:augmentation:setjmp}
The POSIX functions \verb|setjmp| and \verb|longjmp| are special kinds of control flow transfers that allow non-local jumps that bypass the normal call and return sequence. They save and restore native execution context, including the program counter and the registers. In a lifted binary, the relevant context lives partly in emulated state (the threadlocal variables modeling the register file), which a native call to \verb|setjmp| would not capture. We replace calls to these functions with helpers that save and restore the emulated state as well as the native state.

% The replacement introduces one constraint. The POSIX specification states that \verb|longjmp| has undefined behavior if the function containing the matching \verb|setjmp| has already returned. This means that single-block functions will not work, as the \verb|setjmp| block would return to its caller before the \verb|longjmp| can jump back to it. To avoid this, we perform function recovery on any function that contains a \verb|setjmp| in addition to the symbol-based recovery described in \autoref{sec:augmentation:function_recovery}. We also inline the \verb|setjmp| helper into the caller to ensure that \verb|setjmp|'s caller does not prematurely return.

\subsubsection{ABI-Dependent Types}
A few types are encoded differently by \intel{} and \arm{} regardless of which side issues the call. The three we have encountered in our evaluation suite are \verb|long double|, \verb|va_list|, and \verb|struct stat|. For each occurrence, the augmentation pass inserts an explicit conversion at the boundary.

% \tian{Potential cut?}
% \subsection{Retrofitting at the LLVM IR Level}
% Once the IR is well-formed and connected, the LLVM pipeline accepts it
% the way it accepts any other module, and standard or custom passes can
% be applied to add functionality that the \og{} did not have. As an
% example, we have implemented an automatic backward-edge integrity
% check, similar in effect to a shadow stack, as an LLVM pass over the
% lifted IR.

% The lifted binary uses two stacks. The emulated stack holds the \og{}'s
% state, including the return addresses pushed by emulated \verb|call|
% instructions; vulnerabilities in the \og{} that overwrite stack memory
% overwrite the emulated stack. The native stack is the host's own
% stack, used to run the lifted code itself; reaching it requires a
% separate attack on the lifted binary rather than on the \og{}'s logic.

% The pass installs a check at every call site. Before the call, it
% records the expected return address on the native side. After the
% call returns, the basic block following the call site compares the
% emulated program counter to the recorded value; a mismatch aborts the
% program. Figure~\ref{fig:backedge} illustrates the construction.

% The retrofit lives entirely in an LLVM pass; \mirrorball{}'s lifter is
% not modified. Other instrumentation---taint tracking, control-flow
% integrity for forward edges, address sanitizer---can be added through
% the same mechanism.

% --------------------------------------------------------------
\section{Compilation and Linking}
\label{sec:design:compilation_linking}
A typical codebase contains many source files. Each of them is a translation unit that will be translated into an LLVM module. Our pipeline contains a single LLVM module that captures all the superset basic blocks of the input program. To avoid slow compilation and out-of-memory issues, after augmentation, the IR is split into multiple modules before being handed to an unmodified LLVM backend.
%For same-architecture recompilation we target \intel{}; for cross-architecture recompilation we target \arm{}.
We target \intel{} for same architecture recompilation and \arm{} for cross-architecture recompilation.
The backend produces an object file containing the lifted code, which we then link against a small runtime library and a read-only data segment (labeled \verb|.original.binary| in \Cref{fig:architecture}) containing the input binary. The runtime library holds the dispatcher (\verb|dispatch_indirect|, \verb|callback_trampoline_lookup|), the generic external-call trampoline, the \verb|SIGSEGV| handler, and the helpers for \verb|setjmp| and \verb|longjmp|.

The input binary is linked in as data, not as code.
%At no point during execution does the host fetch instructions from the embedded segment; the segment exists so that read-only data accesses from lifted code (e.g., string literals, jump tables, vtables, \cpp{} Run-Time Type Information (RTTI), etc.) resolve to the same bytes they would have resolved to in the input binary. 
At no point during execution does the host fetch instructions from the input binary.
Rather, it is linked so that read-only data accesses from lifted code remain valid (e.g., string literals, jump tables, vtables, \cpp{} Run-Time Type Information (RTTI)). 
%The mechanisms of this embedding are described in \autoref{sec:link_original_binary}.

\begin{table*}[tbp]
\centering
\small
\begin{tabular}{lrrrrrr}
\toprule
\multicolumn{1}{c}{\multirow{2}{*}{\textbf{Benchmark}}} &
  \multicolumn{3}{c}{\textbf{Basic Blocks}} &
  \multicolumn{3}{c}{\textbf{On-Disk Size} (MiB)} \\
  \cmidrule(lr){2-4} \cmidrule(lr){5-7}
\multicolumn{1}{c}{} &
  \multicolumn{1}{c}{\textbf{Source}} &
  \multicolumn{1}{c}{\textbf{Lifted}} &
  \multicolumn{1}{c}{\textbf{Factor}} &
  \multicolumn{1}{c}{\textbf{Source}} &
  \multicolumn{1}{c}{\textbf{Lifted}} &
  \multicolumn{1}{c}{\textbf{Factor}} \\
  \midrule
\astar       &        684 &     49{,}642 &  72.6$\times$ &  0.31 &  32.10 & 103.6$\times$ \\
\bzip2       &        512 &     79{,}556 & 155.4$\times$ &  0.20 &  33.79 & 169.0$\times$ \\
\gcc         & 101{,}676  & 3{,}228{,}499 &  31.8$\times$ &  9.53 & 503.27 &  52.8$\times$ \\
\gobmk       &  22{,}972  &   784{,}931  &  34.2$\times$ &  5.81 & 160.70 &  27.7$\times$ \\
\h264ref     &  12{,}137  &   727{,}091  &  59.9$\times$ &  1.62 &  77.71 &  48.0$\times$ \\
\hmmer       &   3{,}480  &   285{,}186  &  82.0$\times$ &  0.94 &  54.14 &  57.6$\times$ \\
\libquantum  &        882 &     41{,}739 &  47.3$\times$ &  0.14 &  31.66 & 226.1$\times$ \\
\mcf         &        299 &     19{,}556 &  65.4$\times$ &  0.06 &  29.50 & 491.7$\times$ \\
\omnetpp     &  21{,}860  &   421{,}610  &  19.3$\times$ &  3.78 & 112.09 &  29.7$\times$ \\
\perlbench   &  41{,}902  & 1{,}232{,}376 &  29.4$\times$ &  3.04 & 207.28 &  68.2$\times$ \\
\sjeng       &   3{,}644  &   142{,}258  &  39.0$\times$ &  0.35 &  41.85 & 119.6$\times$ \\
\xalancbmk   &  31{,}427  & 2{,}182{,}753 &  69.5$\times$ & 49.52 & 459.94 &   9.3$\times$ \\
\midrule
\textbf{Mean} & & & \textbf{49.6$\times$} & & & \textbf{73.5$\times$} \\
\bottomrule
\end{tabular}
\caption{Cost of brute-force lifting: LLVM basic-block count and on-disk size of the source-compiled \intel{} binary versus the \intel{}-to-\intel{} recompiled binary produced by \mirrorball{}.}
\label{tab:code_expansion}
\end{table*}

\section{Evaluation}

We evaluate \mirrorball{} in four areas: whether the system lifts and recompiles a standard benchmark suite (completeness), what the brute-force strategy costs in IR size (code expansion), what overhead the resulting binaries incur on the original architecture (runtime overhead), and what the system enables when composed with unmodified LLVM backends and instrumentation passes (applications).

\subsection{Methodology}
\label{sec:eval:methodology}
We use the SPECint~2006 suite~\cite{speccpu2006} as our benchmark suite.  SPECint~2006 contains 12 legacy C and C++ programs spanning compilers, interpreters, simulators, and compression workloads, and has been the standard correctness and performance corpus for prior binary lifters and rewriters~\cite{liu2022lifter_sok, altinay2020binrec, wytiwyg, bauman2018superset, janitizer2025, kim2025suri_sound_reassembly,kim2023reassembly_is_hard}. All inputs are binary executables compiled with gcc~13.3.0. Each SPECint program contains one or more workloads; unless otherwise stated, we report numbers for the first workload of each program. 

% Yes I know it's 15 but 21 there's no speed up.
\mirrorball{} produces LLVM~15 bitcode. We recompile lifted bitcode with the flags \texttt{-O2 -max-devirt-iteration=1 -inline-threshold=100}; the lowered optimization level alleviates pressures for some of the LLVM passes that do not scale well with the size expansion.

\intel{} measurements use an AMD EPYC~4564P (4.5\,GHz base clock, 128\,GB DDR5) running Ubuntu~22.04.2. \arm{} measurements use a Neoverse-N1 processor (3.0\,GHz base clock, 64\,GB DDR4) running Ubuntu~22.04.2. We ran each measurement at least three times with \texttt{hyperfine}~\cite{hyperfine} and report means.

\subsection{Completeness}
\mirrorball{} successfully lifted and recompiled all 12 SPECint 2006 programs. Every recompiled binary executed the full reference workload to completion and produced the output matching the source-compiled reference. The same statement holds for both the same-ISA recompilation (\intel{}-to-\intel{}) and cross-recompilation (\intel{}-to-\arm{}) configurations evaluated in the remainder of this section. We make no completeness claim outside the SPECint suite.

\subsection{Code Expansion}
Brute-force disassembly decodes at every byte offset rather than attempting to distinguish code from data, and the resulting superset control-flow graph contains every basic block any execution could possibly reach. The cost of this conservatism shows up in the IR. \Cref{tab:code_expansion} reports the LLVM basic-block count and on-disk size of each binary before and after the same-ISA recompilation.

The block-count expansion ranges from 19.3$\times$ (\omnetpp{}) to 155.4$\times$ (\bzip2{}), with a mean of 49.6$\times$. \bzip2{} is an outlier because the source contains a heavily macro-unrolled loop the compiler turns into a small but dense binary; brute-force decoding inside that region produces many overlapping superset interpretations. The remaining variation across benchmarks tracks properties of the input binary, such as instruction mix, invalid opcode and terminator density, and the proportion of embedded data, rather than any single dominant factor.

On-disk size expands by a larger mean of 74$\times$. Three factors contribute: (i) the original binary is embedded in the lifted output as a read-only data segment for data references, contributing a fixed overhead independent of lifting decisions; (ii) the lifted IR inlines aggressively during recompilation, causing short tail-calling wrappers lifted as single control-flow paths and inlined by the optimizer; and (iii) the brute-force expansion itself contributes the block-count factor. The largest size factors appear for the smallest source binaries (\mcf, \libquantum, and \bzip2); the largest source binary (\xalancbmk) exhibits the smallest size factor (9.3$\times$).

\begin{table}[tbp]
\centering
\small
\begin{tabular}{lrrr}
\toprule
\textbf{Benchmark} & \textbf{Source (s)} & \textbf{Lifted (s)} & \textbf{Slowdown} \\
\midrule
\astar       &  67.02 & 195.50 & 2.92$\times$ \\
\bzip2       &  41.66 & 135.46 & 3.25$\times$ \\
\gcc         &   7.97 &  35.15 & 4.41$\times$ \\
\gobmk       &  23.32 &  86.23 & 3.70$\times$ \\
\h264ref     &  28.09 & 148.83 & 5.30$\times$ \\
\hmmer       &  47.47 & 252.01 & 5.31$\times$ \\
\libquantum  & 100.43 & 514.75 & 5.13$\times$ \\
\mcf         & 137.26 & 259.08 & 1.89$\times$ \\
\omnetpp     & 184.16 & 414.90 & 2.25$\times$ \\
\perlbench   &  68.12 & 329.88 & 4.84$\times$ \\
\sjeng       & 233.61 & 791.66 & 3.39$\times$ \\
\xalancbmk   & 374.39 & 953.27 & 2.55$\times$ \\
\midrule
\textbf{Mean} & & & \textbf{3.74$\times$} \\
\bottomrule
\end{tabular}
\caption{Same-ISA recompilation runtime: source-compiled \intel{} binary versus the same binary lifted to LLVM IR and lowered back to \intel{} by \mirrorball{}.}
\label{tab:x86_overhead}
\end{table}
\subsection{Runtime Overhead}
\label{sec:eval:runtime_overhead}

We measure runtime overhead by lifting each SPECint binary to LLVM IR and lowering it back to an \intel{} executable using the same backend the source compiler would have used. \Cref{tab:x86_overhead} reports the resulting slowdowns. The mean is 3.74$\times$.

To understand where the overhead comes from, we instrumented the lifted binaries with runtime counters at three points the lifter itself introduces: indirect-branch dispatches through the \texttt{dispatch\_indirect} table, calls into external library code through the calling convention adapter, and callbacks from external code back into lifted code through the known-signature path. \Cref{tab:runtime_counters} reports the per-workload totals. The following three observations were made from this data.

\begin{table}[tbp]
\centering
\small
\begin{tabular}{lrrr}
\toprule
\textbf{Benchmark} & \textbf{Indirect} & \textbf{External} & \textbf{Callbacks} \\
\midrule
\astar       &  2{,}780{,}359{,}179 &     7{,}577{,}532 &        0 \\
\bzip2       &              27{,}304 &              488  &        0 \\
\gcc         &     375{,}927{,}871 &     5{,}951{,}985 &  25{,}492 \\
\gobmk       &       31{,}077{,}622 &    78{,}066{,}207 &        0 \\
\h264ref     &  5{,}041{,}322{,}408 &    10{,}682{,}057 &   2{,}153 \\
\hmmer       &       64{,}691{,}682 &   143{,}946{,}279 &        0 \\
\libquantum  &                    0 &    52{,}609{,}064 &        0 \\
\mcf         &                    0 &           470{,}598 &        0 \\
\omnetpp     &  3{,}877{,}621{,}066 & 1{,}683{,}326{,}454 &        0 \\
\perlbench   & 11{,}705{,}591{,}850 &   334{,}395{,}345 &        0 \\
\sjeng       & 21{,}357{,}073{,}580 &           25{,}113 &        0 \\
\xalancbmk   & 10{,}559{,}526{,}324 &   321{,}784{,}160 &        0 \\
\bottomrule
\end{tabular}
\caption{Runtime counts of operations introduced by lifting, collected during one reference-workload run of each round-tripped binary. \emph{Indirect}: dispatches through the indirect-branch table. \emph{External}: calls into native library code through the calling-convention adapter. \emph{Callbacks}: control transfers from external code into lifted code.}% through the known-signature path.}
\label{tab:runtime_counters}
\end{table}

\paragraph{Lifter overhead is present even when dispatch and marshalling activity are minimal.}
\label{sec:eval:lifter_overhead}
Two benchmarks, \mcf{} and \libquantum{}, perform no indirect calls or jumps during execution. Their slowdowns are 1.89$\times$ and 5.13$\times$ respectively. Neither benchmark's overhead can be attributed to indirect-dispatch cost, and \libquantum{}'s external-call rate (530K/s) is moderate compared to the other benchmarks, so we cannot attribute \libquantum{}'s overhead to the external calls either. We attribute the overhead to \mirrorball{}'s register-as-threadlocals memory model: architectural registers are lifted as LLVM threadlocal variables, and the unmodified optimization pipeline cannot promote them to SSA values across basic-block boundaries, so most register-to-register data flow is materialized as load and store traffic to the memory the optimizer cannot prove unaliased. The spread between these two benchmarks indicates that the magnitude of this cost varies substantially with binary structure, but the data does not isolate which properties drive it.

\paragraph{Callbacks are rare in the benchmark suite.}
Callbacks from external code into lifted code are nonzero but not stressed by SPECint. They do not contribute significantly to the overall overhead but the correct implementation of handling callbacks still remains important for completeness. Same applies to \texttt{setjmp}/\texttt{longjmp} handling.

% % TRIM
% \subsection{Application: Backward-Edge Control-Flow Integrity}
% To demonstrate that the lifted IR is amenable to unmodified LLVM
% instrumentation passes, we implemented a backward-edge control-flow
% integrity transformation as a standard LLVM pass operating on the
% lifted module (Section~\ref{sec:shadow_stack}). The pass inserts
% shadow-stack checks at every function-return site in the recovered
% CFG. Because the underlying transformation is an LLVM module pass,
% no part of \mirrorball{} itself was modified to support this
% application.

% Table~\ref{tab:cfi} reports the resulting binary-size overhead.
% The mean size overhead is 16\%. Benchmarks dominated by
% short basic blocks (\h264ref{} at 1.85$\times$, \bzip2{} at
% 1.49$\times$, \hmmer{} at 1.41$\times$) exhibit higher overhead
% because the per-block shadow-stack check is a larger fraction of
% each block's instruction count. Several benchmarks show overhead
% factors below 1.00, reflecting that the inserted instrumentation
% is small enough that downstream optimization and layout choices
% can absorb its space cost. The result we wish to highlight is not
% the overhead figure itself but that the transformation is achieved
% by an unmodified LLVM module pass running on the lifted IR.

\subsection{Application: Static \intel{}-to-\arm{} Cross-Recompilation}
\label{sec:eval:cross_isa}

To demonstrate that the lifted IR can be directly consumed by the unmodified LLVM \arm{} backend, we lifted each SPECint~2006 binary from \intel{} to LLVM IR using \mirrorball{}, then lowered the lifted IR to \arm{} using the same unmodified LLVM \arm{} backend that lowers the source-compiled \arm{} binaries. The resulting \arm{} executable is self-contained: no runtime translation support is required on the host machine.

As shown in \Cref{tab:x86_to_arm}, across the nine benchmarks the LLVM \arm{} backend can compile as a single module, the mean slowdown is 6.64$\times$. Three benchmarks (\gcc{}, \omnetpp{}, and \xalancbmk{}) produce large code sections that LLVM struggles to compile as a single module. We thus split those binaries into submodules using \texttt{llvm::SplitModule} and lose optimization opportunities across the split boundaries. Including these three benchmarks raises the mean to 12.33$\times$. The split-module overhead is therefore not a fundamental cost of cross-recompilation but a consequence of a known scaling limitation, addressable by either incremental LLVM improvements, emitting lifted code as multiple modules from the start, or by splitting the modules in a way that preserves more optimization opportunities.

\begin{table}[tbp]
\small
\centering
\begin{tabular}{lrrr}
\toprule
\multicolumn{1}{c}{\textbf{Benchmark}} &
\multicolumn{1}{c}{\textbf{Source (s)}} &
\multicolumn{1}{c}{\textbf{Lifted (s)}} &
\multicolumn{1}{c}{\textbf{Factor}} \\
\midrule
\astar              & 140.91 &    507.85 &  3.60$\times$ \\
\bzip2              &  83.47 &    749.99 &  8.99$\times$ \\
\gobmk              &  52.72 &    819.11 & 15.54$\times$ \\
\h264ref            &  65.67 &    434.69 &  6.62$\times$ \\
\hmmer              &  91.16 &    640.86 &  7.03$\times$ \\
\libquantum         & 275.86 & 1{,}924.64 &  6.98$\times$ \\
\mcf                & 419.89 & 1{,}335.59 &  3.18$\times$ \\
\perlbench          & 173.98 &    538.34 &  3.09$\times$ \\
\sjeng              & 478.63 & 2{,}256.36 &  4.71$\times$ \\
\midrule
\textbf{Mean (9 benchmarks)} & & & \textbf{6.64$\times$} \\
\midrule
\gcc{} \dag         &  21.36 &     764.72 & 35.80$\times$ \\
\omnetpp{} \dag     & 347.22 & 16{,}462.03 & 47.41$\times$ \\
\xalancbmk{} \dag   & 587.63 &  2{,}971.15 &  5.06$\times$ \\
\midrule
\textbf{Mean (12 benchmarks)} & & & \textbf{12.33$\times$} \\
\bottomrule
\end{tabular}
\caption{Cross-recompilation runtime. Source-compiled \arm{} versus \intel{}-to-\arm{} cross-recompiled, both run on identical \arm{} hardware. Programs marked \dag{} produce lifted IR that LLVM struggles to compile as a single module, so these were compiled as split modules at the cost of optimization.}
\label{tab:x86_to_arm}
\end{table}

\omnetpp{} exhibits the highest cross-recompilation slowdown and also the highest number of external calls. This suggests that external calls contribute significantly to the overhead. The role of external calls is sharper in this configuration than in the same-ISA recompilation, because the extra stack management required to address the mismatch between the different number of argument registers on \intel{} and \arm{} (see \Cref{sec:augmentation:external_stack_management}).

%A 7-fold (or, including the module-split benchmarks, 12-fold) is not negligible, and we do not minimize it. The argument we make is narrower: this is the first system, to our knowledge, that produces a fully static, self-contained recompilation of non-trivial binaries using only unmodified LLVM components on the backend.

\section{Discussion and Future Work}

\paragraph{Positioning.}
We do not think brute-force lifting, as built here, is the right engineering choice for a production deployment today. Despite our best efforts to improve the performance, the overheads in \Cref{sec:eval:runtime_overhead,sec:eval:cross_isa} are large. And for cross-ISA migration specifically, they are worse than what mature dynamic binary translators already deliver on hardware that exists now. We present this as the central empirical finding of the paper. Building a fully static, heuristics-free lifter that requires no runtime translation support on the target was, prior to this work, an open question. We show that it is possible, and we measure precisely what it costs to do it this conservatively. Our measurements serve as a reference baseline: a maximally conservative baseline against which a future system that recovers a superset CFG more selectively, for example by narrowing it with a soundly verified static analysis rather than an unsound heuristic, can measure how much of this cost it recovers. Whether such a system would still satisfy the fully static, no-runtime-fallback property we start from here is a right next question for this line of research.

\paragraph{Register model.}
The runtime overhead reported in \Cref{sec:eval:lifter_overhead} is a consequence of the register-as-threadlocal memory model. Addressing this would allow the unmodified LLVM pipeline to promote the register state to SSA values. This represents the single largest performance opportunity for the implementation.

\paragraph{Self-modifying code.}
\label{sec:future_work:self_modifying_code}
\mirrorball{} does not accept binaries that generate code at runtime. This is a permanent limitation of the approach: self-modifying code is fundamentally incompatible with fully static lifting, since the runtime generated code is not known without executing the program. Supporting it would require shipping the lifting machinery with the recompiled binary, which directly contradicts the fully static claim.

\paragraph{Broader input scope.}
Three further classes of binary input are excluded from the prototype but are possible extensions rather than fundamental limitations. Multithreaded binaries have been handled by prior recompilation systems~\cite{rocha2022lasagne, deshpande2024polynima}. \cpp{} exception unwinding requires parsing the \verb|.eh_frame| section and regenerating equivalent metadata for the lifted IR; prior recompilers~\cite{wytiwyg, deshpande2024polynima, smithson2013secondwrite, remill} have likewise deferred this, but nothing in the brute-force approach prohibits it. Extending the front end to non-\intel{} input ISAs is also engineering rather than research, requiring a new lifting front end and revised calling-convention marshalling at external call boundaries.

\section{Conclusion}

This paper introduces \mirrorball{}, a fully static binary lifting system built with \emph{brute-force lifting}. Rather than deciding which bytes of an x86-64 ELF binary represent code and which represent data, \mirrorball{} treats every byte offset as a potential instruction boundary and every reachable address as a potential branch target. The resulting superset control flow graph conservatively contains every feasible control-flow path, including paths hidden behind overlapping instruction encodings. Statically unresolvable indirect branches reduce to lookups in a dispatch table over the superset CFG, removing the need for any translation machinery to accompany the lifted program on the target.

The output is standard LLVM IR accepted by the unmodified LLVM pipeline. As a result, lifted binaries can be processed by the same retargetable backends, optimization passes, and instrumentation passes that LLVM already provides for source code. We demonstrate this through fully static cross-recompilation of \intel{} binaries to \arm{}, achieved by feeding the lifted IR directly to LLVM's \arm{} backend without modifying any component of the toolchain.

The cost, as our evaluation shows, is substantial: the lifting time and runtime overhead inherent to producing and executing a conservative superset CFG, much of which is never reached, range from several-fold to two orders of magnitude depending on the metric (\Cref{sec:eval:runtime_overhead,sec:eval:cross_isa}). These costs are large enough that we do not expect brute-force lifting, in the form presented here, to be the right engineering choice for a production deployment: mature dynamic binary translators already deliver cross-ISA execution at a fraction of the overhead we report, on hardware that exists today. This is the paper's central empirical finding. \mirrorball{} shows that a fully static, heuristics-free lifter with no runtime translation support on the target is achievable at all, and quantifies precisely what that guarantee costs. We offer this as a reference point for the field: a maximally conservative baseline against which future systems that relax the brute-force strategy, for instance by narrowing the superset CFG with a soundly verified static analysis rather than an unsound heuristic, can measure how much of this cost they recover.

% \begin{acks}
% The acknowledgments go here.
% \end{acks}

\bibliographystyle{ACM-Reference-Format}
\bibliography{bib/mirrorball}

\appendix

\end{document}